\documentclass[preprintnumbers,article,amsmath,amssymb,floatfix,10pt,prd,twocolumn,
superscriptaddress,nofootinbib]{revtex4}

\usepackage{bbm}
\usepackage{amsfonts}
\usepackage{mathrsfs}
\usepackage{latexsym}
\usepackage{epsfig}
\usepackage{epstopdf}
\usepackage{epstopdf}
\usepackage{graphicx}
\usepackage{amssymb}
\usepackage{amsmath}
\usepackage{dcolumn}
\usepackage{bm}
\usepackage{color}
\usepackage{comment}
\usepackage{xcolor}
\def\be{\begin{equation}}
\def\ee{\end{equation}}
\def\beq{\begin{eqnarray}}
\def\eeq{\end{eqnarray}}

\usepackage[latin1]{inputenc}
\usepackage{graphicx, psfrag}
\usepackage[colorlinks=true, citecolor=magenta, urlcolor = magenta, linkcolor= cyan, bookmarks=true]{hyperref}
\usepackage{float}
\usepackage{amsfonts}
\usepackage{hyperref}
\usepackage{subfigure}
\usepackage{booktabs}
\usepackage{orcidlink}
\usepackage{enumitem}

\begin{document}
\title{ Possible wormholes in $f(R)$ gravity sourced by solitonic quantum wave and cold dark matter halos and their repulsive gravity effect}

\author{Abdelghani Errehymy\orcidlink{0000-0002-0253-3578}}
\email[Email:]{abdelghani.errehymy@gmail.com}
\affiliation{Astrophysics Research Centre, School of Mathematics, Statistics and Computer Science, University of KwaZulu-Natal, Private Bag X54001, Durban 4000, South Africa}

\author{Youssef Khedif}
\email[Email: ]{youssef.khedif@gmail.com}
\affiliation{Laboratory of High Energy and Condensed Matter Physics, Department of Physics, Faculty of Sciences A\"in Chock, University Hassan II, P.O. Box 5366 Maarif Casablanca 20100, Morocco}

\author{Orhan Donmez}
\email[Email: ]{orhan.donmez@aum.edu.kw}
\affiliation{College of Engineering and Technology, American University of the Middle East, Egaila 54200, Kuwait}

\author{Mohammed Daoud}
\email[Email:]{m$_{}$daoud@hotmail.com}
\affiliation{Department of Physics, Faculty of Sciences, Ibn Tofail University, P.O. Box 133, Kenitra 14000, Morocco}
\affiliation{Abdus Salam International Centre for Theoretical Physics, Miramare, Trieste 34151, Italy}

\author{Kairat Myrzakulov}
\email[Email: ]{krmyrzakulov@gmail.com}
\affiliation{Department of General and Theoretical Physics, L.N. Gumilyov Eurasian National University, Astana 010008, Kazakhstan }

\author{Sabit Bekov}
\email[Email: ]{ss.bekov@gmail.com}
\affiliation{Department of General and Theoretical Physics, L.N. Gumilyov Eurasian National University, Astana 010008, Kazakhstan }
\affiliation{Kozybayev University,  Petropavlovsk 150000, Kazakhstan}

\begin{abstract}%
In this paper, we present new generalized wormhole (WH) solutions within the context of $f(R)$ gravity. Specifically, we focus on $f(R)$ gravitational theories formulated in the metric formalism, with our investigation centered on a power-law form represented by $f(R) = \epsilon R^{\chi}$. Here, $\epsilon$ is an arbitrary constant, and $\chi$ is a real number. Notably, this form possesses the advantageous property of reducing to Einstein gravity when $\epsilon=1$ and $\chi=1$. To obtain these novel WH solutions, we establish the general field equations for any $f(R)$ theory within the framework of Morris-Thorne spacetime, assuming metric coefficients that are independent of time. By utilizing an anisotropic matter source and a specific type of energy density associated with solitonic quantum wave (SQW) and cold dark matter (CDM) halos, we calculate two distinct WH solutions. We thoroughly investigate the properties of the exotic matter (ExoM) residing within the WH geometry and analyze the matter contents through energy conditions (ECs). Both analytical and graphical methods are employed in this analysis to examine the validity of different regions. Notably, the calculated shape functions for the WH geometry satisfy the necessary conditions in both scenarios, emphasizing their reliability. Our investigations into specific parameter ranges in both scenarios revealed the presence of ExoM. This ExoM is characterized by an energy-momentum tensor that violates the null energy condition (NEC) and, consequently, the weak energy condition as well, in the vicinity of the WH throats. Furthermore, we investigated the repulsive effect of gravity and discovered that its presence results in a negative deflection angle for photons following null geodesics. Importantly, we observed that the deflection angle consistently exhibits negative values across all $r_0$ values in both scenarios, indicating the manifestation of the repulsive gravity effect. Finally, we compare the obtained WH solutions utilizing both distributions, as well as the $f(R)$ power-law-like models, in order to assess the feasibility of energetic configurations for WHs within SQW and CDM systems.
\end{abstract}

\maketitle

\date{\today}


\maketitle

\section{Introduction}\label{sec1}
The observed data strongly supports the $\Lambda$CDM model of standard cosmology, which indicates that the composition of our universe is estimated to be approximately 67.4 \% dark energy (DE), 29.6 \% dark matter (DM), and 4 \% baryonic matter \cite{Planck:2018vyg}. DM plays a significant role in the formation, evolution, and merging of galaxies and their clusters, primarily through its gravitational interaction \cite{Trujillo-Gomez:2010jbn}. { The existence of DM in galaxies was first asserted by Zwicky \cite{Zwicky:1937zza}, who used the virial theorem to make this observation. It is believed that spiral galaxies have Universal Rotation Curves (URC), and the existence of DM in the galactic halos can be confirmed by its gravitational pull on the URC \cite{Roberts:1978zza, Roszkowski:2017nbc}. Based on the Navarro-Frenk-White (NFW) density profile which characterizes the distribution of CDM halos, and the observed flat rotation curves of galaxies, the authors \cite{Rahaman:2013xoa, Rahaman:2014pba} showed that galactic halos can support the existence of traversable WHs. Following this work, galactic halo WHs have been discussed in different theories of gravity. Sharif \textit{et al.} \cite{Sharif:2014bsa, Sharif:2016bvi} have explored galactic halo WH solutions in various modified theories of gravity, taking into account the NFW density profile. The formation of traversable WHs in DM isothermal galactic halos has also been discussed by Rahaman \textit{et al.} \cite{Rahaman:2016hkk, Sarkar:2019uhk}, considering both the NFW and URC density profiles. Islam \textit{et al.} \cite{Islam:2018ciy} examined WH formation in the Dragonfly 44 galaxy, considering two different DM profiles --- the ultra-diffuse galaxy King's model and the generalized NFW model. They found that only the generalized NFW model can support traversable WHs. Furthermore, the formation of spherically symmetric traversable WHs in DM halos with isotropic pressure has been discussed by Xu \textit{et al.} \cite{Xu:2020wfm}. More recently, the evolution of topologically deformed WHs in DM halos has been studied in \cite{Ovgun:2018uin}. Finally, the interesting phenomenon of gravitational lensing by galactic halo WHs has been investigated by Kuhfittig \cite{Kuhfittig:2013hva}. 

An alternative model that has received significant attention is the fuzzy DM scenario \cite{Hu:2000ke}. The fuzzy DM model proposes that DM is composed of axion-like particles with a mass of approximately $m\approx 10^{-22}$ eV \cite{Goodman:2000tg,Schive:2014dra,Schive:2014hza,Hui:2016ltb,Bernal:2017oih}. This extremely low mass corresponds to a de Broglie wavelength that is as large as the typical length scale of a galaxy. In this model, the observed DM cores within galaxies are hypothesized to be soliton cores. These soliton cores arise from a balance between the quantum pressure due to the uncertainty principle and gravity. Importantly, the observed properties of these DM cores would directly reflect the mass ($m$) of the fuzzy DM particles. Given the fuzzy DM particle mass $m$, the characteristics of the DM cores can be fully determined by solving the coupled Schr\"{o}dinger-Poisson equation \cite{Widrow:1993qq}. This equation governs the behavior of the quantum wave function of the fuzzy DM particles and their gravitational interactions. Several research groups have compared the predicted fuzzy DM core structure with observational data. For example, Deng \textit{et al.} \cite{Deng:2018jjz} examined a broad class of theoretical light scalar DM models. These models are governed by a potential $V$ and assume a complex scalar field with a global $U(1)$ symmetry. They demonstrated that within the analytical framework of their model, a single axion mass cannot simultaneously explain the observed wide range of DM core radii ($r_s$) and reproduce the observed core-scaling relation $\rho_s \approx r_s^{-1}$ \cite{Burkert:2015vla,Kormendy:2014ova,Rodrigues:2017vto}. However, this interesting result does not take into account the formation of DM halos through the process of cosmic structure formation. The first self-consistent $3D$ cosmological simulations of fuzzy DM halo formation were presented by Schive \textit{et al.} \cite{Schive:2014dra, Schive:2014hza}. These simulations confirmed that all DM halos develop a distinct, gravitationally self-bound solitonic core with a universal core density distribution.} 

Traversable WHs are hypothetical objects predicted by the theory of general relativity (GR). They are conceptualized as tunnels in spacetime that have the potential to connect distant regions within the same universe or even different universes \cite{Morris:1988cz,Visser:1995cc}. Recent works have explored this subject in a more fundamental context \cite{Maldacena:2013xja,Maldacena:2017axo,Maldacena:2020sxe}, and its applications have extended to condensed matter systems as well \cite{Gonzalez:2009je,Alencar:2021ejd}. Typically, traversable WHs require ExoM as their energy source, but in modified theories of gravity, non-ExoM can serve as the source for the WH geometry \cite{Moti,Alencar,Sadeghi:2022sto}. Over the past few years, there has been considerable interest in the field of $f(R)$ gravity. In this approach, the traditional Einstein's field equations are extended by replacing the Ricci scalar curvature, denoted as $R$, with a general function $f(R)$ in the gravitational action. One notable advantage of $f(R)$ gravity, in contrast to the standard $\Lambda$CDM cosmology based on  GR, is that it eliminates the need for DE or the introduction of new matter fields to explain cosmic inflation \cite{Kehagias:2013mya, Ketov, Aziz:2021evx, Sharma:2022tce} and the current accelerated expansion of the universe \cite{Nojiri, Sotiriou}. A particularly significant $f(R)$ model, proposed by Starobinsky \cite{Starobinsky:1979ty, Starobinsky:1980te}, includes an additional quadratic term in the Einstein-Hilbert action that involves the Ricci scalar. This model, characterized by only one free parameter, predicts a de Sitter phase in which the quadratic term dominates. Importantly, this model is in agreement with experimental data obtained from the Planck satellite \cite{Planck:2018jri}.

Significant progress has been made in the study of WHs\footnote{For more detailed information, the readers can refer to the comprehensive study presented in \cite{Bronnikov:1973fh, Bronnikov:2002rn, Boehmer:2012uyw, Capozziello:2012hr, Bronnikov:2016xvj, Bronnikov:2017kvq, Bronnikov:2018uje, Bronnikov:2019gsr, Bronnikov:2020ikh, DeFalco:2021klh, Bronnikov:2021xao, Bolokhov:2021fil, Bronnikov:2023saq, Errehymy:2024yey, Errehymy:2023rsm, Errehymy:2023rnd, Errehymy:2024cgy}. These works delve into the intricacies of WH solutions within the context of modified gravity, providing a valuable resource for further exploration.}, although direct evidence has not yet been obtained. One notable exploration is the concept of WHs existing within galactic halos, which has been investigated in \cite{Rahaman:2013xoa, Rahaman:2014pba}. The observed flat galactic rotation curve can be consistent with a traversable WH geometry in the spacetime of a galactic halo, as indicated by research. Investigations have also been conducted to explore the possibility of detecting these WHs through gravitational lensing \cite{Kuhfittig:2013hva}, with further analysis provided in \cite{Nandi:2006ds}. Notably, supermassive black holes (BHs) located at the centers of galaxies \cite{Errehymy:2023xpc, Mandal:2022oma, Paul:2023vys, Myrzakulov:2023rkr} are considered as potential WH candidates that originated from the early Universe. To differentiate between BHs and WHs, a proposed method involves identifying orbiting hot spots \cite{Li:2014coa}, in addition to examining their Einstein-ring systems \cite{Tsukamoto:2012xs}. Gravitational lensing is a remarkable phenomenon first explored by Einstein \cite{Einstein:1936llh} as part of his groundbreaking work in GR. This effect occurs when an extremely massive object, such as a galaxy or BHs, bends the path of light passing nearby, much like a lens focusing light. This light distortion allows astronomers to study the nature and properties of the lensing object and the background light source \cite{Dyson:1920cwa, Eddington:1920cwa}. The first observational confirmation of gravitational lensing came in 1979, when astronomers detected a double quasar image caused by the bending of light \cite{Walsh:1979nx}. Since this pioneering discovery, gravitational lensing has become an indispensable tool for probing the universe beyond our solar system. It has enabled the detection of exoplanets, DM, and even the elusive DE \cite{Sahu:2017ksz}. One intriguing aspect of gravitational lensing is the possibility of ``\textit{eternal}'' light rings --- stable orbits where light can be bent and trapped indefinitely around certain cosmic objects like BHs \cite{Virbhadra:1999nm, Bozza:2002zj}. The study of these strong and weak lensing effects has provided information about the nature of gravity and the structure of the universe \cite{Tsukamoto:2016qro, Jusufi:2017mav, Javed:2019qyg}. As our observational capabilities continue to advance, gravitational lensing remains a crucial tool for unlocking the mysteries of the universe.

Within recent studies, researchers have made significant progress in obtaining distinct electromagnetic signatures from thin accretion disks surrounding static spherically symmetric traversable WHs \cite{Bambi:2013jda, Harko:2009xf}. The revealed findings provide an exciting prospect for astrophysical observations to differentiate and distinguish between different geometries of traversable WHs by analyzing the emitted spectra. Moreover, there have been further suggestions and proposals for detecting astrophysical indicators associated with traversable WHs, which can be explored in \cite{Safonova:2001vz, Cramer:1994qj}. As the endeavor to detect traversable WHs progresses, it becomes increasingly important to understand their composition in terms of matter and geometry. This serves as the central objective of the present article, which aims to compile and present comprehensive predictions within the framework of the $f(R)$ theory of gravity. The article seeks to contribute to our understanding of the intricate relationship between matter and geometry in WHs, facilitating advancements in their detection and characterization. Motivated by these arguments, our investigation delves into the existence of generalized traversable WHs within the framework of $f(R)$ gravity. These WHs are assumed to be supported by SQW and CDM distributions. To do this, we will exploit the general field equations of any $f(R)$ theory that were previously developed through the metric formalism. Subsequently, we will focus on a specific power-law $f(R)$ model that describes modified gravity. Next, we will calculate the shape functions taking into account the influence of the SQW and CDM distributions. Moreover, we will investigate the deflection angle of photons traveling along null geodesics. Finally, we will compare the WH solutions obtained with the aforementioned SQW and CDM distributions, as well as the $f(R)$ power-law-like models, in terms of the feasibility of energetic configurations for WHs within SQW and CDM systems.

This paper is structured as follows: In Sec. \ref{sec2}, we provide a concise overview of the $f(R)$ gravity theory and examine the geometry of a WH within the context of a static and spherically symmetric spacetime. We also explore the constraints imposed on the metric functions of the WH geometry and discuss their implications on the matter distribution threading the spacetime. In Sec. \ref{sec3}, we analyze the ECs from a global perspective. In Sec. \ref{sec4}, we present a novel approach to WHs sourced by SQW and CDM halos in $f(R)$ gravity. Furthermore, in Sec. \ref{sec5}, we investigate the repulsive behavior of gravity. In Sec. \ref{sec6}, we interpret and analyze the constructed WHs sourced by SQW and CDM halos, as well as the effect of repulsive gravity in the framework of $f(R)$ gravity. Finally, in Sec. \ref{sec7}, we provide concluding remarks summarizing the key findings of this study. 

\section{An overview of $f(R)$ gravity: A brief background}\label{sec2}
The action for $f(R)$ gravity can be rewritten in the following form
\begin{small}
\begin{eqnarray}\label{Eq1}
S &=& \int d^{4}x \sqrt{-g}\left(\frac{1}{16\pi}f(R) + \mathcal{L}_{m}(g_{\mu\nu},\psi_{m})\right),
\end{eqnarray}
\end{small}
where $S$ is the action, $d^4x$ represents the spacetime volume element, $g$ is the determinant of the metric tensor $g_{\mu\nu}$, $R$ is the Ricci scalar, $f(R)$ is the modified gravity function, and $\mathcal{L}_m(g_{\mu\nu},\psi_m)$ is the matter Lagrangian density involving the matter fields $\psi_m$ interacting with the metric tensor $g_{\mu\nu}$.

The field equation resulting from varying the action with respect to the metric tensor can be rewritten as
\begin{small}
\begin{eqnarray}\label{Eq2}
F(R)R_{\mu\nu}-\frac{1}{2}g_{\mu\nu}f(R)+\left[g_{\mu\nu}\Box-\nabla_{\mu}\nabla_{\nu}\right]F(R)=8\pi T_{\mu\nu},
\end{eqnarray}
\end{small}
where $F(R)\equiv df(R)/dR$ represents the derivative of the function $f(R)$ with respect to $R$. This equation establishes a relationship between the Ricci tensor $R_{\mu\nu}$, the curvature scalar $R$, the function $f(R)$, the derivative of $f(R)$ denoted as $F(R)$, and the energy-momentum tensor $T_{\mu\nu}$ which describes the matter and energy distribution. On the left-hand side, the equation includes terms involving $F(R)$, $R_{\mu\nu}$, $g_{\mu\nu}$, the d'Alembertian operator represented by $\Box=g^{\mu\nu}\nabla_{\mu}\nabla_{\nu}$, and the covariant derivatives $\nabla_{\mu}$ acting on $F(R)$. These terms represent the gravitational field. On the right-hand side, $T_{\mu\nu}$ represents the energy-momentum tensor associated with matter fields. It describes the matter content and serves as the source of the gravitational field described by the left-hand side. The energy-momentum tensor can be expressed as
\begin{eqnarray}\label{Eq3}
T_{\mu\nu}=(\rho+p_{t})u_{\mu}u_{\nu}-p_{t}g_{\mu\nu}+(p_{r}-p_{t})V_{\mu}V_{\nu},
\end{eqnarray}
where $\rho$, $p_r$, $p_t$ denote the energy density and the radial and transverse stresses, respectively. The four-velocity is denoted by $u_{\mu}$, and $V_{\mu}$ is a radial four-vector; they satisfy $u^{\mu}u_{\mu}=1$ and $V^{\mu}V_{\mu}=-1$. 

Now, to construct a traversable WH solution, we consider the line element proposed by Morris and Thorne \cite{Morris:1988cz}, which can be written as
\begin{eqnarray}\label{Eq5}
ds^2= -e^{2\hat{\nu}(r)}dt^2 + e^{2\hat{\mu}(r)} dr^2 +r^2 d\Omega^2.
\end{eqnarray}
Here, $ d\Omega^2= d\theta^2 + \text{sin}^2 \theta d\phi^2 $, $e^{2\hat{\mu}(r)} = \left[1-\frac{b(r)}{r}\right]^{-1}$, while $\hat{\nu}(r)$ and $b(r)$ are the redshift function and shape function, respectively. The construction of a traversable WH solution requires the fulfillment of the following conditions:
\begin{enumerate}
	\item The WH connects two asymptotically flat regions at the throat, where the throat radius is characterized by a global minimum $r=r_0$. The radial coordinate ranges from $r_0$ to infinity, i.e., $r \in [r_0, \infty)$.
	\item The redshift function $\hat{\nu}(r)$ must indeed remain finite at all points along the traversable WH, ensuring its regularity throughout and guaranteeing the absence of horizons and singularities. Hence, $e^{2\hat{\nu}(r)}>0$ for all $r>r_0$.
	In the case of an ultrastatic WH, where there is no gravitational acceleration, $ \hat{\nu}(r) $ is set to zero, resulting in $ e^{2\hat{\nu}(r)}=1 $. This implies that a particle dropped from rest remains at rest \cite{Morris:1988cz,Cataldo:2017ard}.
	\item The flaring-out condition is given by $\frac{b(r)-rb'(r)}{b^2(r)}>0$, and must hold at or near the throat, i.e., $r=r_0$.
	\item The conditions $b(r_0)=r_0$ and $b'(r_0)\leq 1$ are necessary for all $r\geq r_0$, with $b'(r_0)=1$ at the throat. Furthermore, for $r>r_0$, it is required that $b(r)<r$.
	\item Asymptotic flatness is ensured by $\hat{\nu}(r)\rightarrow 0$ and $b(r)/r \rightarrow 0$ as $r\rightarrow \infty$.
\end{enumerate}
 These conditions govern the behavior of the redshift and shape function in the WH geometry\footnote{Readers are referred to \cite{Morris:1988cz}, for more detailed information.}. It's worth noting that meeting these essential properties is necessary but not sufficient to ensure the physical viability of a WH solution. While satisfying the conditions mentioned earlier is important, additional considerations must be taken into account. These include the satisfaction of ECs, stability analysis, and the behavior of matter fields. These factors play a crucial role in constructing physically realistic WHs. 

To acquire the expressions for the energy density ($\rho$), radial pressure ($p_r$), and tangential pressure ($p_t$), it is imperative to formulate specific equations based on the pertinent field equations associated with the problem. The exact methodologies for deriving these quantities hinge upon the subsequent formulas
\begin{small}
\begin{eqnarray}\label{Eq16}
\rho&=&\frac{df}{dR}\frac{b'(r)}{r^{2}},\\
p_{r}&=&-\frac{df}{dR}\frac{b(r)}{r^{3}}
+\frac{d^{2}f}{dR^{2}}\frac{rb'(r)-b(r)}{2r^{2}}-\frac{d^{3}f}{dR^{3}}\left(1-\frac{b(r)}{r}\right),~~\\
p_{t}&=&-\frac{1}{r}\frac{d^{2}f}{dR^{2}}
\left(1-\frac{b(r)}{r}\right)+\frac{1}{2r^{3}}\frac{df}{dR}\left(b(r)-rb'(r)\right).~~~~~~~\label{Eq18}
\end{eqnarray}
\end{small}
The relationship between the Ricci scalar $R$ and the shape function in the case of WH geometry can be expressed by $R = \frac{2b'(r)}{r}$, where $b'(r)$ represents the derivative of the shape function with respect to $r$. 
In our study focusing on the generation of a WH solution using SQW and CDM halos in $f(R)$ gravity, we will consider a specific form of the function $f(R)$ within the power-law $f(R)$ gravity model. This specific form is given by
\begin{eqnarray}\label{Eq4}
     f(R) = \epsilon R^\chi.
\end{eqnarray}
Here $\epsilon$ is an arbitrary constant and $\chi$ is the power law exponent. This power-law $f(R)$ gravity model has undergone extensive scrutiny in relation to late-time acceleration in the cosmos \cite{Jaime:2012yi}. Nevertheless, comprehensive solar system examinations have conclusively invalidated its viability as a plausible substitute for elucidating fundamental phenomena such as DE, matter dominance, radiation dominance, and unified inflation \cite{Nojiri:2006gh, Nojiri:2003ft, Nojiri:2007as, Nojiri:2007cq, Cognola:2007zu}.

In the extensive literature, scientists have extensively examined various specific scenarios within the power-law $f(R)$ gravity model, as evidenced by studies \cite{Capozziello:2003tk, Carloni:2004kp} such as $\chi=-1$ and $\chi=3/2$. Notably, the case where $\chi=3/2$ is of significant interest due to its intriguing feature of conformal equivalence to Liouville field theory. Carloni and his colleagues performed an in-depth analysis of the cosmic dynamics associated with the power-law $f(R)$ model, as extensively documented in their study \cite{Carloni:2004kp}. However, despite its limitations, it continues to be widely utilized as a simplified model for theoretical investigations and serves as a valuable tool for exploring various aspects of modified gravity \cite{Faraoni:2011pm, Capozziello:2006dp, Borka:2012tj, Zakharov:2014kka}. The power-law $f(R)$ model was employed in another study by Faraoni \textit{et al.} \cite{Faraoni:2011pm} to investigate the solar system's chameleon process. The results showed that $\chi$ needs to be somewhat near to $1$ in order to be consistent. Therefore, this study suggests that the power-law $f(R)$ gravity model is not a viable candidate for replacing DE in a realistic context. The work presented by Bronnikov \textit{et al.} in \cite{Bronnikov:2006pt,Bronnikov:2010tt} establishes an important no-go theorem regarding WH solutions in $f(R)$ gravity theories. The authors demonstrate that static and spherically symmetric WHs cannot exist in these theories without the presence of ExoM that violates the NEC condition. This finding has significant implications. While it may be possible to construct a localized throat region without relying on ExoM, the authors conclude that a genuine, globally configured WH does not appear feasible under reasonable and generic conditions within the context of scalar-tensor and $f(R)$ gravity frameworks. In other words, the no-go theorem puts strict constraints on the viability of WH solutions in these modified gravity theories, unless one is willing to invoke the presence of exotic, NEC-violating matter, which remains a highly speculative and problematic proposition. This work highlights the challenges faced in realizing traversable WH geometries from first principles in these alternative gravity models. The principal aim here is to assess the viability of the power-law $f(R)$ gravity model in producing physically reasonable WH solutions. By examining the behavior of WH within this specific model, we seek to understand the constraints and unique features associated with $f(R)$ gravity theories. The findings of this study will broaden the implications of modified gravity theories, elucidating the dynamics of matter and energy within such theories and expanding our understanding of the nature of spacetime. 
\begin{figure*}
\centering
\includegraphics[width=8.5cm,height=5.45cm]{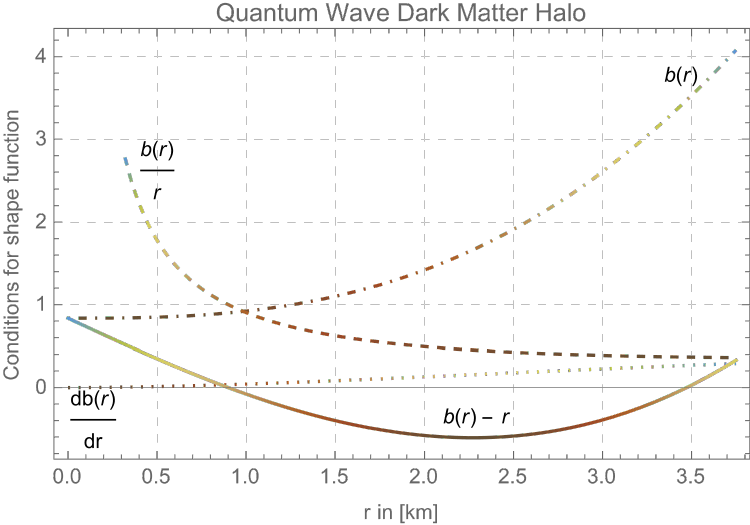}
\includegraphics[width=8.5cm,height=5.45cm]{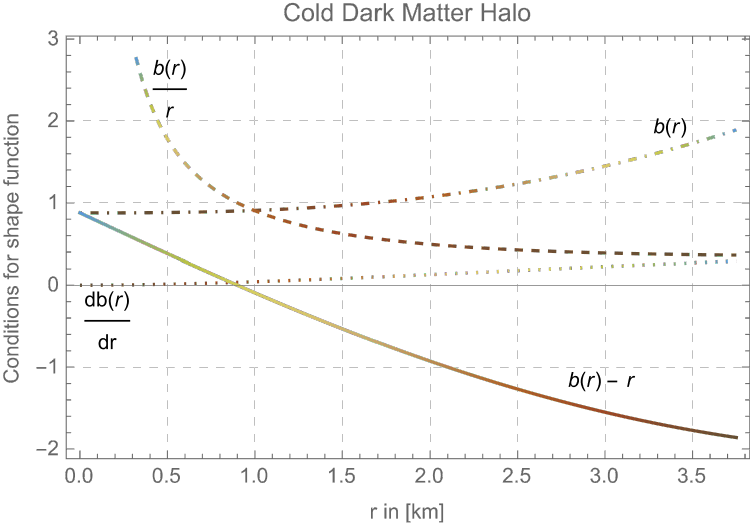}
\caption{ { Graphical representation of the shape functions for two distinct scenarios: the SQW halo--(left panel) and the CDM halo--(right panel). Here, the throat radius is uniformly set to $r_0 = 0.9$.}}\label{fig1}
\end{figure*}
\section{Analyzing ECs: A global perspective}\label{sec3}
Moving on to the constraints on the metric functions of the WH geometry, these constraints are dictated by the Einstein field equations and provide information about the matter threading the spacetime. In the context of the WH geometry, the metric functions $\hat{\nu}(r)$ and $b(r)$ are intimately connected to the mass-energy density, radial pressure, and tangential pressure of the matter. The curvature of spacetime and the distribution of matter and energy are related, according to Einstein's field equations. Therefore, violations of ECs may arise from the particular form of the metric functions that satisfy the fundamental features previously mentioned. The energy density and elements of the matter's energy-momentum tensor are constrained by ECs, which are inequalities\footnote{For additional details on these conditions, we refer readers to consult the following Refs. \cite{Raychaudhuri:1957jwp, Hawking:1973uf, Visser:1995cc, Morris:1988cz, Hochberg:1998ii, Hochberg:1998ha, Capozziello:2014bqa, Kontou:2020bta} for further information.}. By placing constraints on the matter content, these requirements are essential in ensuring the physical plausibility of the spacetime solutions. Four ECs---the null energy condition (NEC), weak energy condition (WEC), dominant energy condition (DEC), and strong energy condition (SEC)---will be looked at in this study. These conditions are expressed as follows
{\begin{small}
 \begin{eqnarray}
NEC&:&~\rho+ p_{i} \geq 0,\\
WEC&:&~\rho \geq0,~ \rho+ p_{i}\geq 0,\\
DEC&:&~\rho- |p_{i}| \geq 0,\\
SEC&:&~ \rho+ \overset{3}{\underset{i=1}{\Sigma}} p_{i}\geq 0,~ \rho+ p_{i}\geq 0,
 \end{eqnarray}
\end{small}
where $i =$ space index}. The ECs discussed above play a vital role in the existence of WH solutions. They indicate the necessity of ExoM for the formation of the throat radius in the WH geometry. Studying the geometry of WHs and the associated ECs is crucial for understanding the theoretical and physical implications of these intriguing spacetime structures and the possibility of their existence in the universe.

\begin{figure*}
\centering
\includegraphics[width=8.5cm,height=5.45cm]{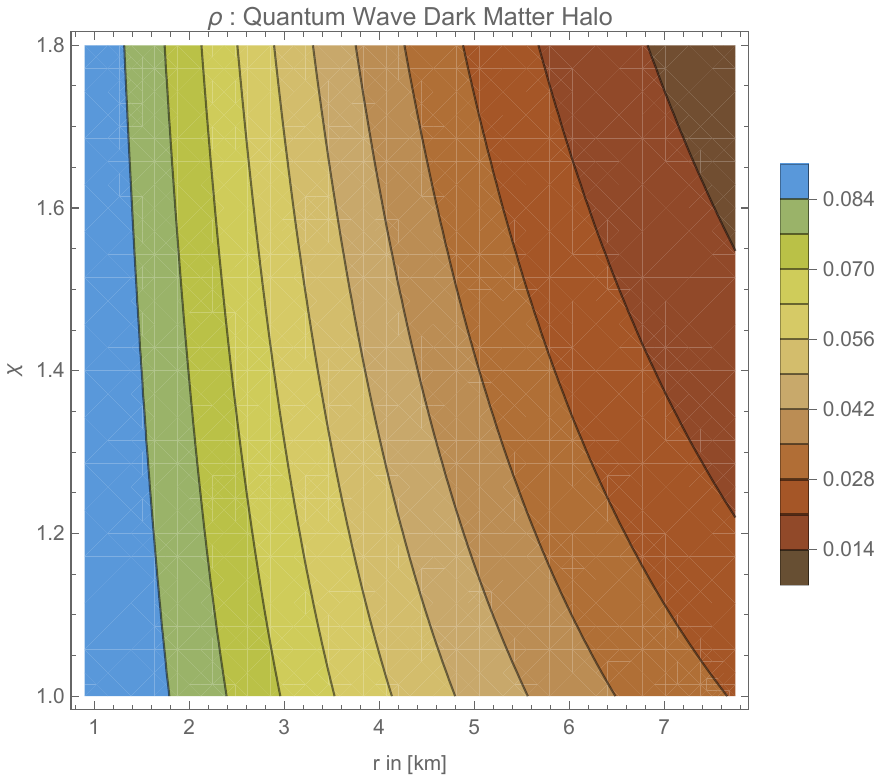}
\includegraphics[width=8.5cm,height=5.45cm]{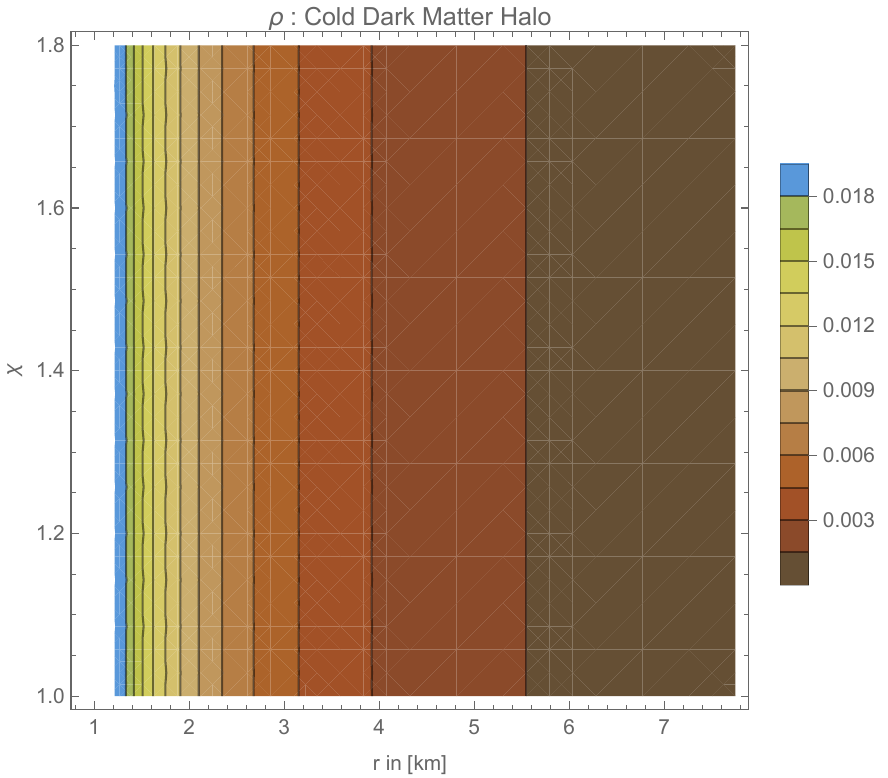}
\caption{ { The behavior of matter density, represented by $\rho$, for both considered scenarios. Here, the throat radius is uniformly set to $r_0 = 0.9$. The specific parameter values used in these plots are as follows: $\alpha =0.9$, $\rho_s = 0.095$ in Planck units, and $r_s = 4.7$  in Planck units.}}\label{fig4}
\end{figure*}

\begin{figure*}
\centering
\includegraphics[width=8.5cm,height=5.45cm]{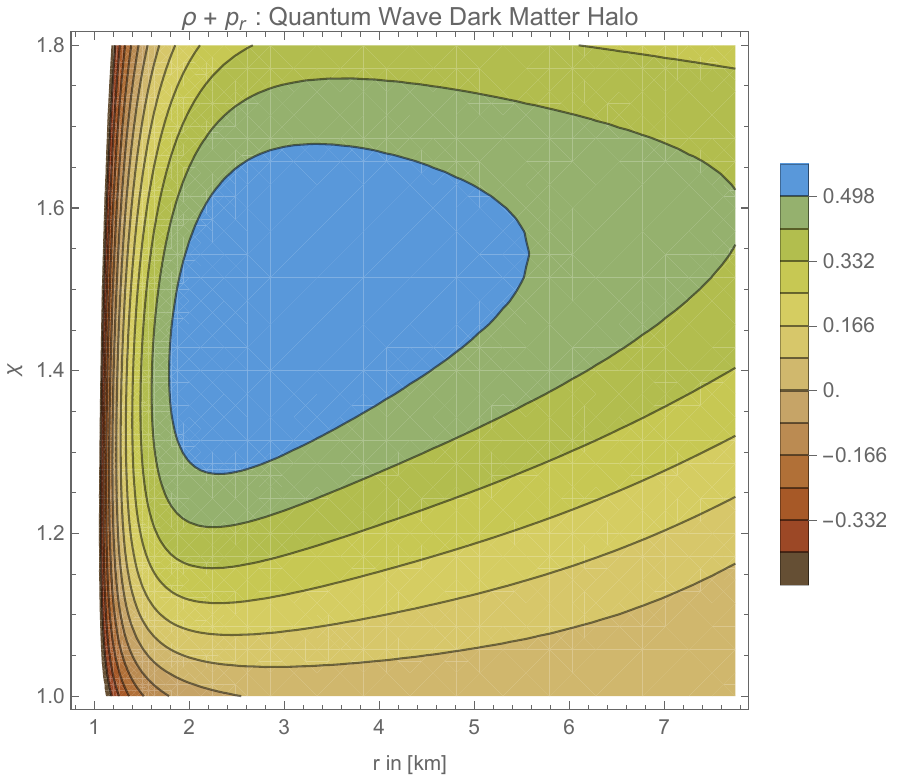}
\includegraphics[width=8.5cm,height=5.45cm]{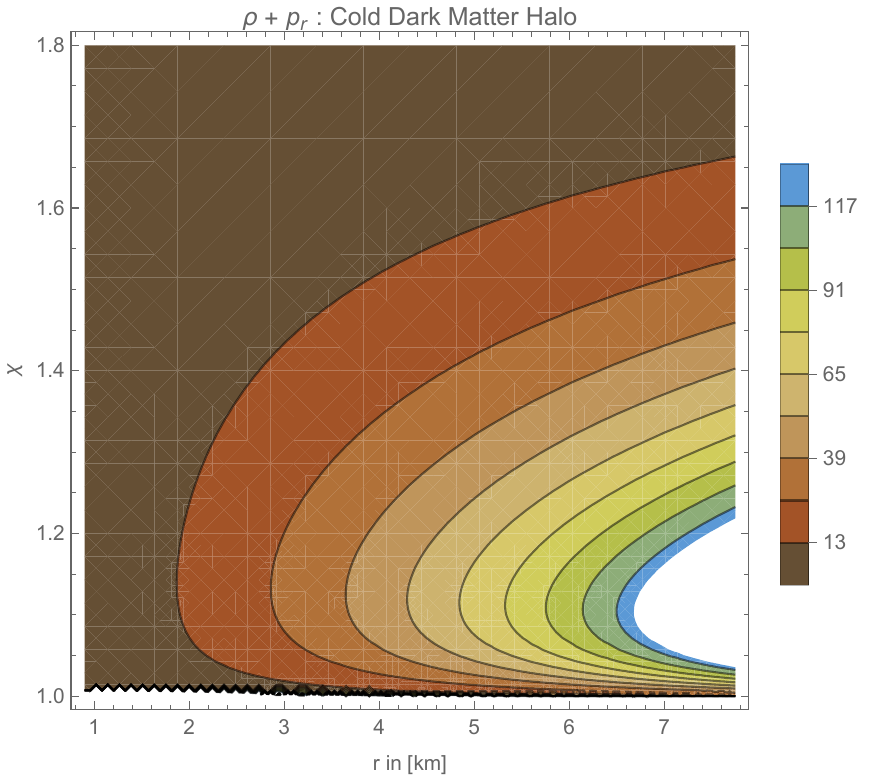}
\caption{ { The behavior of NEC and WEC, represented by $\rho+p_r$, for both considered scenarios. The specific parameter values used in these plots are the same as those in Fig. \ref{fig4}.}}\label{fig5}
\end{figure*}

\begin{figure*}
\centering
\includegraphics[width=8.5cm,height=5.45cm]{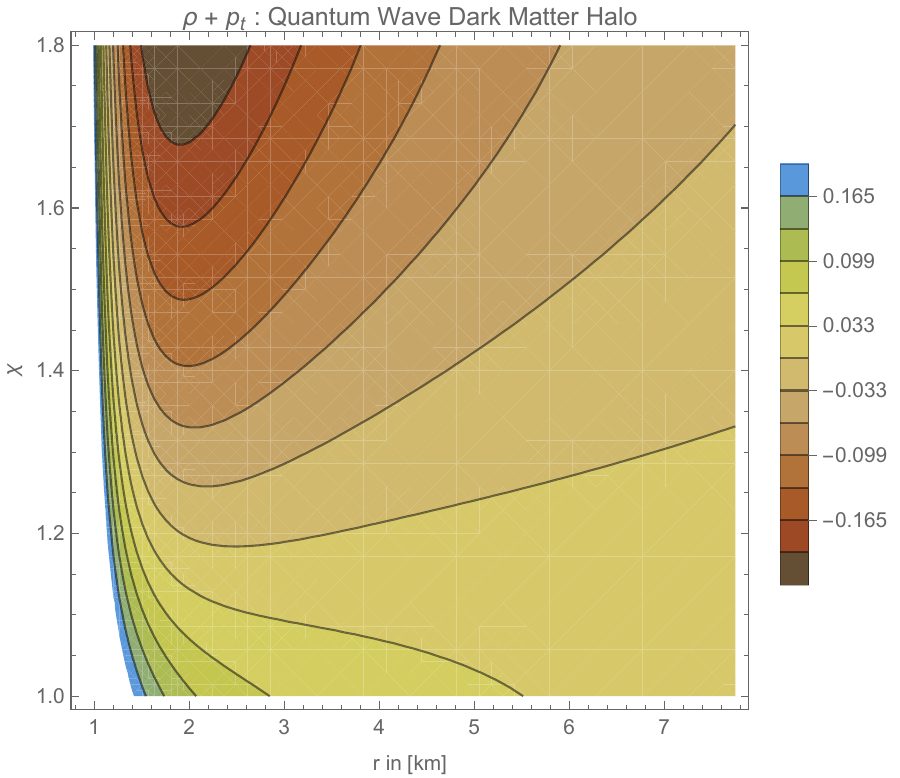}
\includegraphics[width=8.5cm,height=5.45cm]{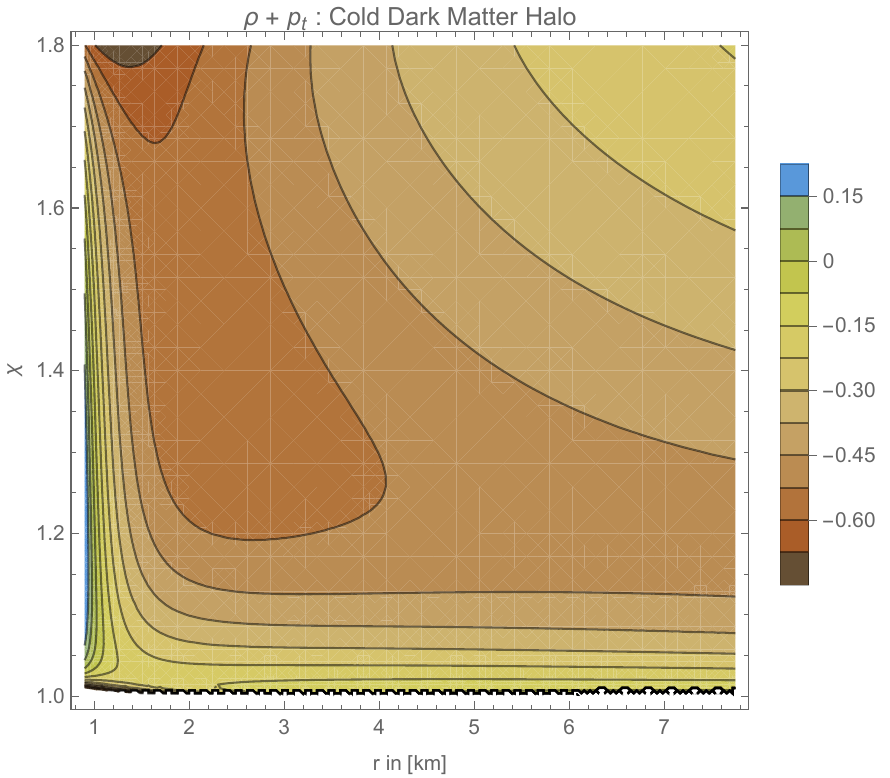}
\caption{ { The behavior of NEC and WEC, represented by $\rho+p_t$, for both considered scenarios. The specific parameter values used in these plots are the same as those in Fig. \ref{fig4}.}}\label{fig6}
\end{figure*}

\begin{figure*}
\centering
\includegraphics[width=8.5cm,height=5.45cm]{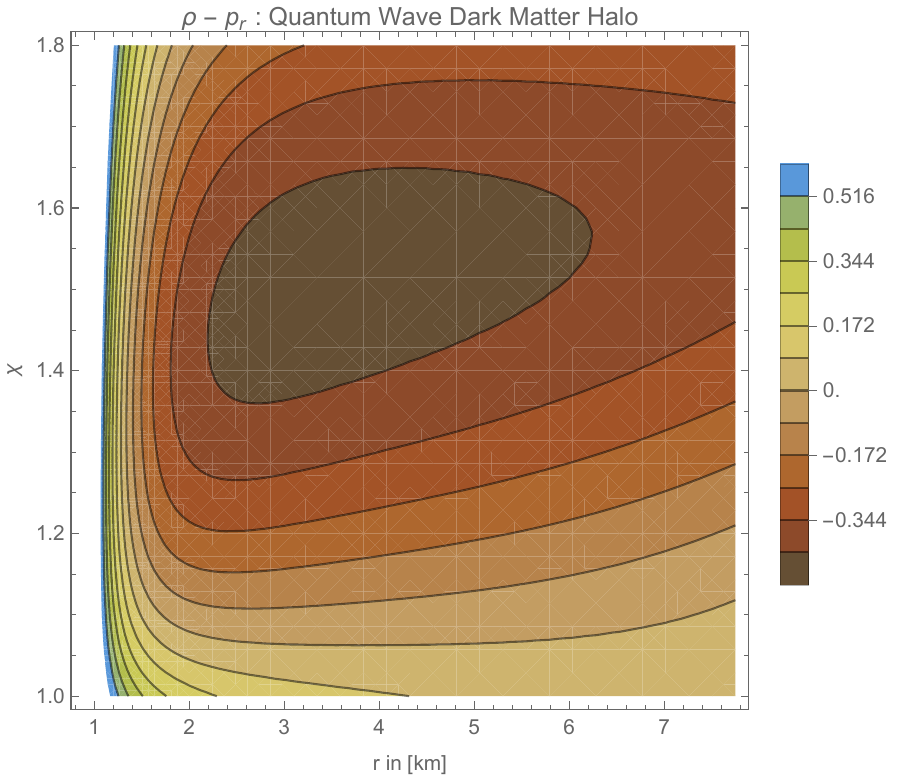}
\includegraphics[width=8.5cm,height=5.45cm]{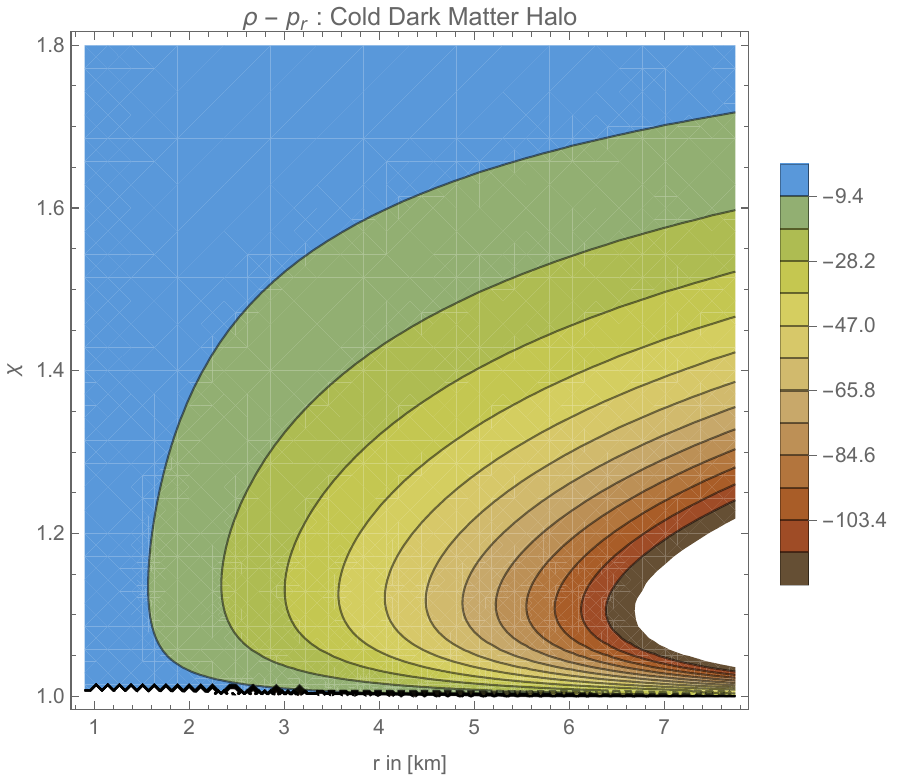}
\caption{ { The behavior of DEC, represented by $\rho-p_r$, for both considered scenarios. The specific parameter values used in these plots are the same as those in Fig. \ref{fig4}.}}\label{fig7}
\end{figure*}

\begin{figure*}
\centering
\includegraphics[width=8.5cm,height=5.45cm]{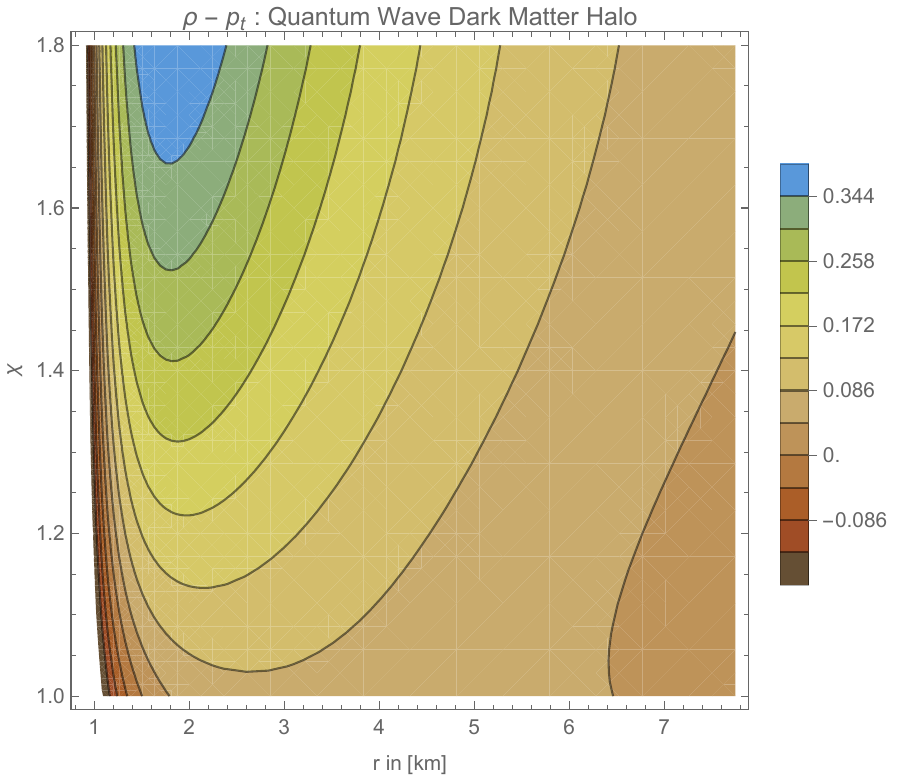}
\includegraphics[width=8.5cm,height=5.45cm]{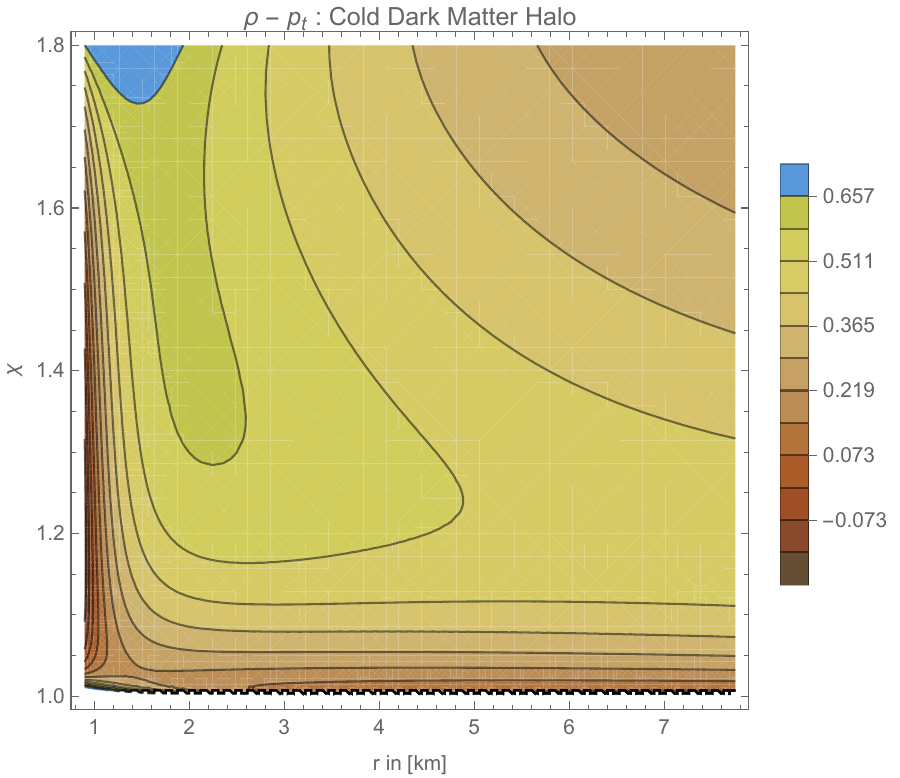}
\caption{ { The behavior of DEC, represented by $\rho-p_t$, for both considered scenarios. The specific parameter values used in these plots are the same as those in Fig. \ref{fig4}.}}\label{fig8}
\end{figure*}

\begin{figure*}
\centering
\includegraphics[width=8.5cm,height=5.45cm]{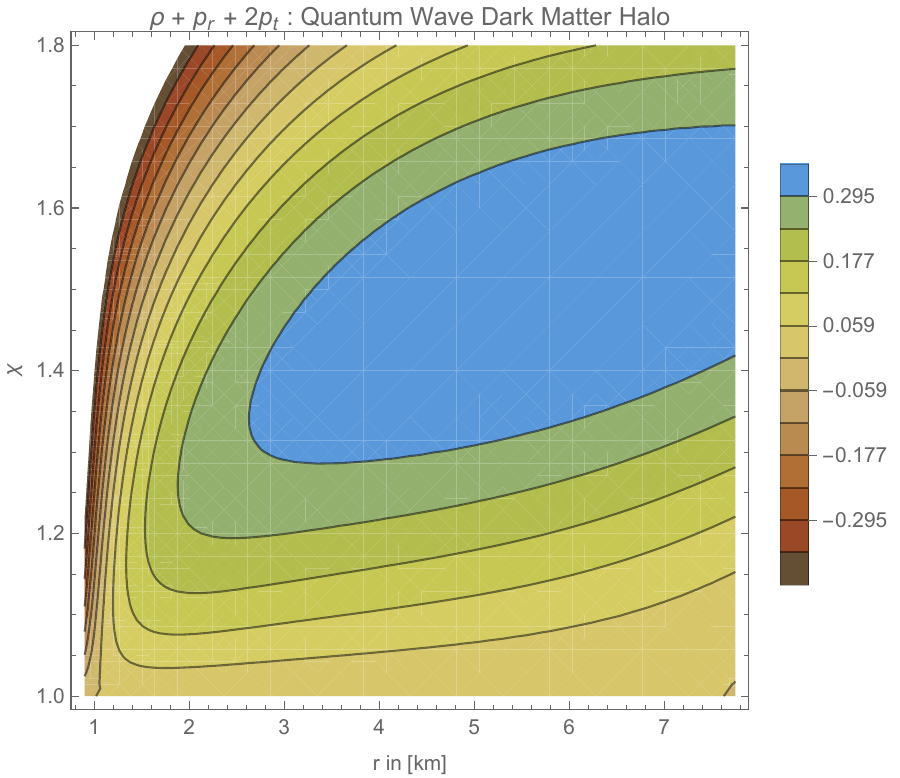}
\includegraphics[width=8.5cm,height=5.45cm]{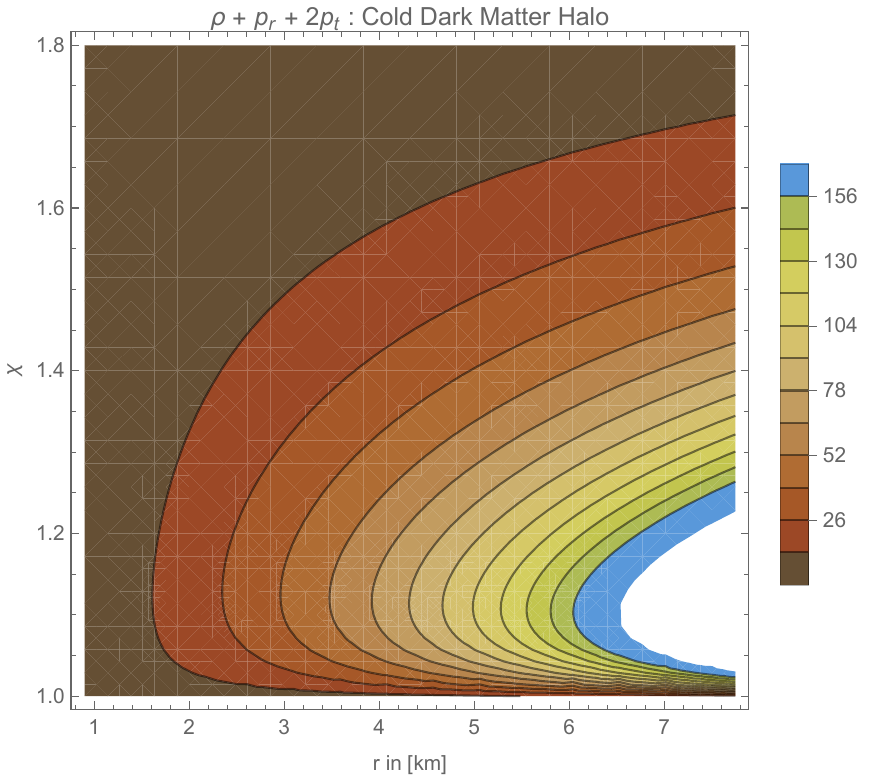}
\caption{ { The behavior of SEC, represented by $\rho+p_r+2p_t$, for both considered scenarios. The specific parameter values used in these plots are the same as those in Fig. \ref{fig4}.}}\label{fig9}
\end{figure*}

\begin{figure*}
\centering
\includegraphics[width=8.5cm,height=5.45cm]{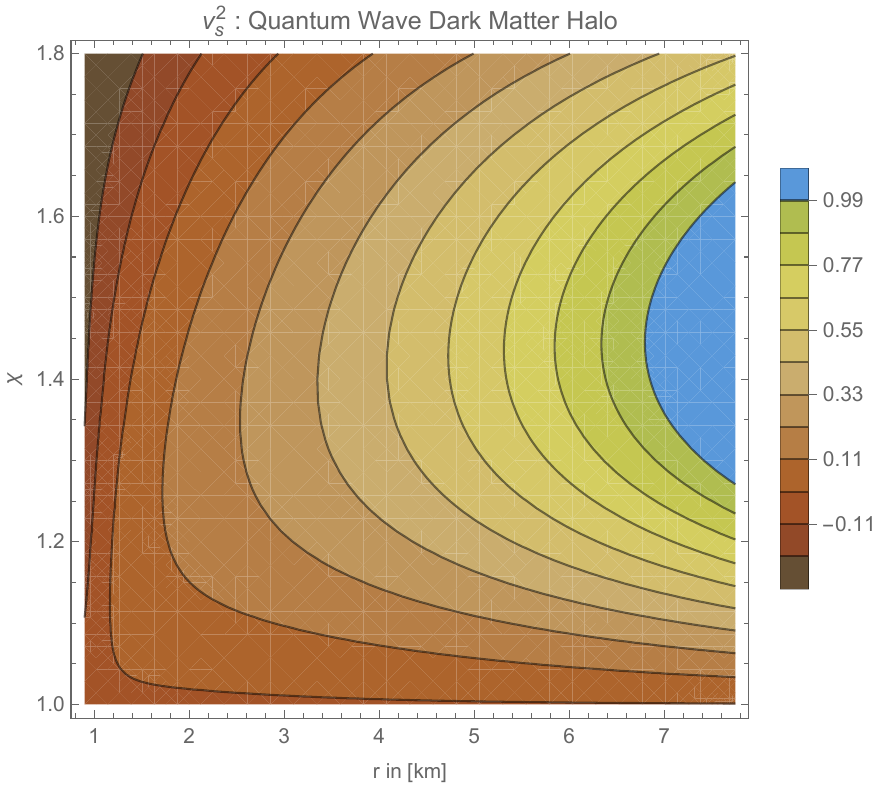}
\includegraphics[width=8.5cm,height=5.45cm]{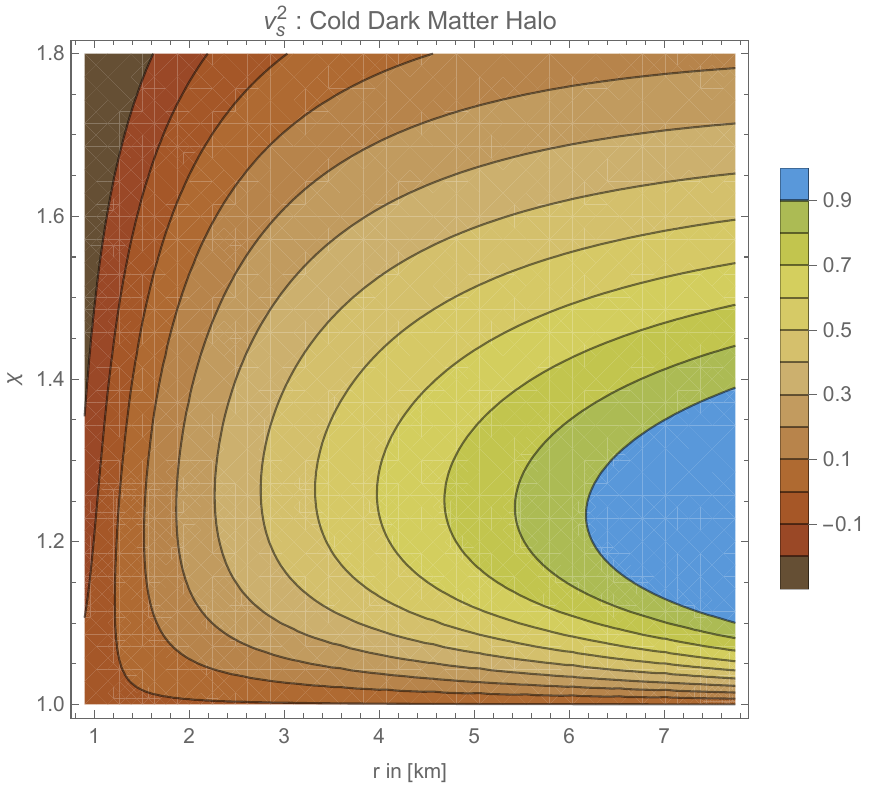}
\caption{{ The behavior of the adiabatic sound speed, represented by $v_s^2$, for both considered scenarios. The specific parameter values used in these plots are the same as those in Fig. \ref{fig4}.}}\label{fig11}
\end{figure*}

\begin{figure*}
\centering
\includegraphics[width=8.5cm,height=5.45cm]{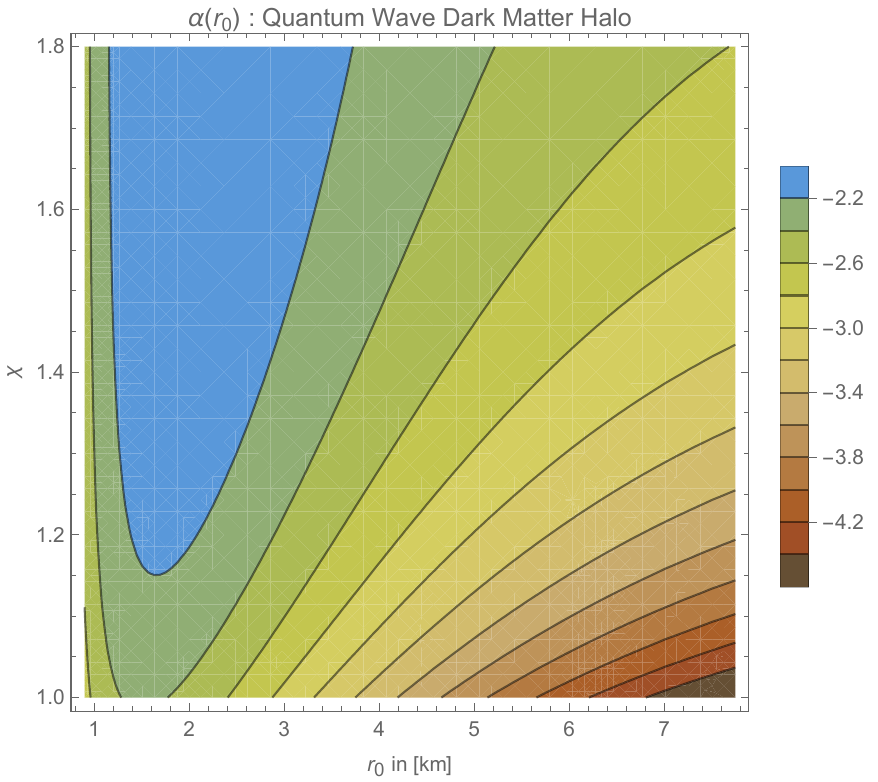}
\includegraphics[width=8.5cm,height=5.45cm]{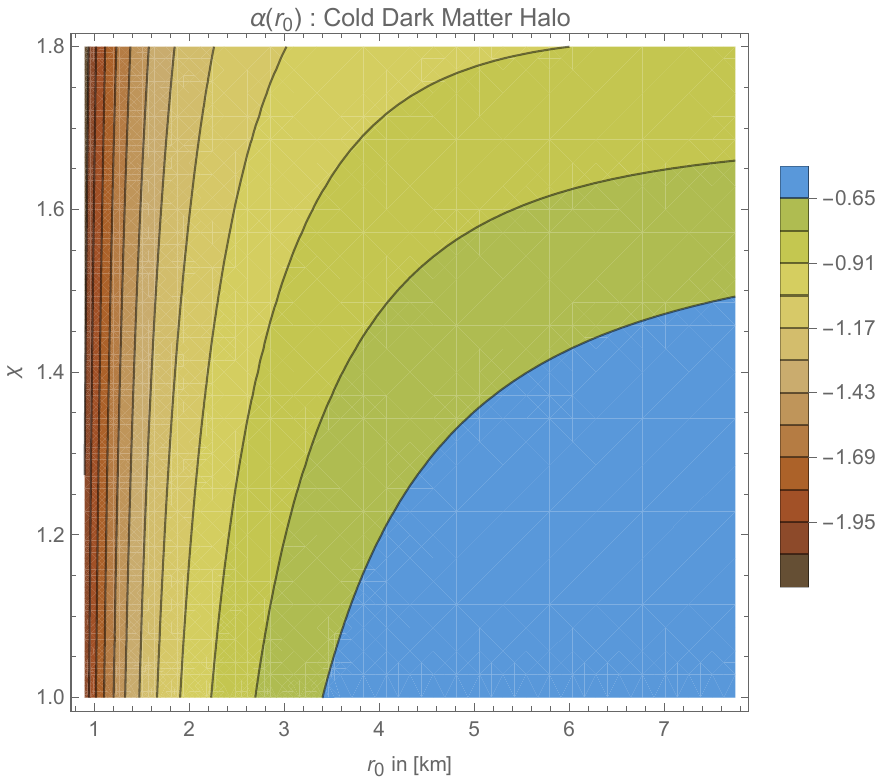}
\caption{{ The behavior of the numerical results of the deflection angle of photons, represented by $\alpha(r_0)$, for both considered scenarios. The specific parameter values used in these plots are the same as those in Fig. \ref{fig4}.}}\label{fig10}
\end{figure*}

\section{WHs sourced by SQW and CDM halos in $f(R)$ gravity: A novel approach}\label{sec4}

In this section, we will derive new WH solutions in the background of $f(R)$ gravity, considering two different assumptions for the material content: SQW and CDM halos. We will utilize the field equations of $f(R)$ gravity to calculate the shape function, taking into consideration the effects of the SQW and CDM distributions. 

\subsection{SQW Halo}
In 2014, Schive \textit{et al.}~\cite{Schive:2014dra} introduced a novel soliton matter density distribution, commonly known as the ``solitonic core". This distribution is mathematically expressed as
\begin{equation} \label{e1}
    \rho(r) = \rho_s\left[1+\alpha\left( \frac{r}{r_s}\right)^2 \right]^{-8}.
\end{equation}
The soliton core density, $\rho_s$, and soliton core radius, $r_s$, were defined by Herrera-Mart\'in \textit{et al.} \cite{Herrera-Martin:2017cux}. The matter distribution discussed above was extensively examined in the seminal study by Schive \textit{et al.} \cite{Schive:2014hza}. The specific radius, $r_s$, is predicted to be the half-density radius, which can be precisely calculated as the constant $\alpha = \sqrt[8]{2}-1 \approx 0.09051$.  Furthermore, in Eq. \eqref{e1}, the value of $\rho_s$ is given as
\begin{equation} \label{e2}
    \rho_s = 2.4\text{x}10^{12} m_\varrho^{-2}  \left[\frac{r_s}{\text{pc}}\right]^{-4}\frac{M_\odot}{\text{pc}^{3}},~~~\text{where}~~~m_\varrho = \frac{m_\text{b}}{10^{-22}\text{eV}},
\end{equation}
with $m_\text{b}$ representing the boson mass parameter. Herrera-Mart\'in \textit{et al.}~\cite{Herrera-Martin:2017cux} argued that the boson mass is a crucial parameter that can be hypothesized to have a single and universal value for all galaxies in the universe.

For a more generic and comparative study, we introduce the parameter $n$ to replace the exponent $8$ in Eq. \eqref{e1}. Thus, the revised form of the solitonic core can be expressed as
\begin{equation} \label{e6}
\rho(r) = \rho_s\left[1+\alpha\left( \frac{r}{r_s}\right)^2 \right]^{-n}.
\end{equation}
In our current analysis, we aim to derive a new WHs sourced by a SQW distribution using Eq. \eqref{e6}, while considering the influence of the boson mass parameter. It is important to note that for exploring specific cosmic scenarios, only whole-number integers are typically assigned to $n$. While it may be theoretically possible to select a value of $n$ other than $8$, it may not be consistent or compatible with recent observational evidence in a physical sense. Furthermore, for the $M87$ galaxy, the estimated soliton core radius is approximately $r_s = 156~\text{pc}$.

\subsubsection{Shape Function Based on SQW Halo}
Here, we aim to determine the WH solution by employing the concept of a shape function, which involves the utilization of a SQW distribution. To achieve this, we will directly apply the field equations of $f(R)$ gravity and compute the shape function, considering the impact of the SQW distribution. By substituting Eq. (\ref{e6}) into Eq.(\ref{Eq16}), we can derive the following expression for the shape function, allowing us to calculate the exact form of the shape function for the SQW halo
\begin{small}
\begin{eqnarray}
 b(r) =\frac{1}{(2 n+1) r_s^2} L_{r} \, X_{r}  Y_{r}^{1/n} + \mathcal{C}_a,
\end{eqnarray}
\end{small}
where $\mathcal{C}_a$ is the integrating constant. By imposing the condition $b(r_0) = r_0$, we can determine the integration constant $\mathcal{C}_a$, leading to the final form of the shape function for the SQW halo
\begin{widetext}
\begin{small}
\begin{eqnarray}\label{sh1}
 b(r) = r_0 + \frac{1}{(2 n+1) r_s^2}\left[ L_{r} \, X_{r}  Y_{r}^{1/n} -  L_{r_0} \, X_{r_0}  Y_{r_0}^{1/n}\right], 
\end{eqnarray}
\end{small}
where
\begin{small}
\begin{eqnarray*}
L_{r} &=& n r \left(\alpha  r^2+r_s^2\right),
 X_{r} = _2F_1\left(1,1+\frac{1}{2 n};2+\frac{1}{2 n};-\frac{r^2 \alpha }{r_s^2}\right),~
 Y_{r} = \frac{2^{1-n} \rho_s}{\alpha  n} r^{n+1} \left(\frac{\alpha  r^2}{r_s^2}+1\right)^{-n},\\
 L_{r_0} &=& n r_0 \left(\alpha  r_0^2+r_s^2\right),~
 X_{r_0} = _2F_1\left(1,1+\frac{1}{2 n};2+\frac{1}{2 n};-\frac{r_0^2 \alpha }{r_s^2}\right),~
 Y_{r_0} = \frac{2^{1-n} \rho_s}{\alpha  n} r_0^{n+1} \left(\frac{\alpha  r_0^2}{r_s^2}+1\right)^{-n}.
\end{eqnarray*}
\end{small}

Building upon the assumptions mentioned earlier, we can express the remaining two nonvanishing components of the $f(R)$ gravity field equations as follows
\begin{small}
\begin{eqnarray}\label{eeq24}
p_r&=& \frac{\alpha  2^{n-3} n  Y_{r}^{-3/n}}{(2 n+1) r^2 r_s^2} \frac{ Y_{r}}{r^n} \Bigg[L_{r_0} \, X_{r_0} \left(-\left(n^2-3 n+2\right) r^4+(n-1) r^2  Y_{r}^{1/n}+4  Y_{r}^{2/n}\right)  Y_{r_0}^{1/n}-L_{r} \, X_{r} \Big(-\left(n^2-3 n+2\right) r^4+(n-1)  \nonumber\\&\times&r^2 Y_{r}^{1/n}+4  Y_{r}^{2/n}\Big)  Y_{r}^{1/n}+(2 n+1) r_s^2 \Big(-\left(n^2-3 n+2\right) r^5 +\left(n^2-3 n+2\right) r^4 r_0-\left((n-1) r^2 r_0  Y_{r}^{1/n}\right)\nonumber\\&&-4 r_0  Y_{r}^{2/n}+(n-1) r^3  Y_{r}^{2/n}\Big)\Big],\label{eeq25}\\
p_t &=& \frac{\alpha  2^{n-2} n  Y_{r}^{-2/n}}{(2 n+1) r^2 r_s^2}
 \frac{ Y_{r}}{r^n} \Big[-L_{r_0} \, X_{r_0} \left((n-1) r^2+ Y_{r}^{1/n}\right)  Y_{r_0}^{1/n}+L_{r} \, X_{r} \Big((n-1) r^2 + Y_{r}^{1/n}\Big)  Y_{r}^{1/n} -(2 n+1) r_s^2 \Big((n-1) r^3\nonumber\\&+&r^2 (r_0-n r_0)-r_0  Y_{r}^{1/n}+r  Y_{r}^{2/n}\Big)\Big].\label{eeq26}
\end{eqnarray}
\end{small}
\end{widetext}
We can also readily investigate the ECs using the equations presented in (\ref {e6}), (\ref {eeq24}), and (\ref {eeq26}) for further quantitative assessment.

\subsection{CDM Halo}
Once again, the NFW profile \cite{Navarro:1996gj}, which characterizes the distribution of CDM halo, was originally derived by analyzing $N$-body systems. Mathematically, it can be expressed as
\begin{small}
\begin{equation}\label{e7}
\rho(r) = \frac{\rho_s}{(r/r_s)(1+r/r_s)^2},
\end{equation}
\end{small}
where $r_s$ is the typical radius and $\rho_s$ is the density at zero radius of the universe. For the $M87$ galaxy, the specific parameters associated with the NFW profile are as follows: The characteristic density is given by $\rho_s = 0.008 \times 10^{7.5}\text{M}_{\odot}/ \text{kpc}^3$ (as stated in \cite{Oldham:2016}), and the scale radius is ${r_s} = 130\text{kpc}$ (also reported in \cite{Oldham:2016}).

\subsubsection{Shape Function Based on CDM Halo}
Once again, we will proceed with calculating the WH solution using a shape function that incorporates a CDM halo. This calculation will be carried out within the context of $f(R)$ gravity. In this subsection, we will apply Eq. (\ref{e7}) to obtain a new WH solution that is influenced by the distribution of the CDM halo. To accomplish this, we will directly utilize the field equations of $f(R)$ gravity and determine the shape function while considering the impact of the CDM halo distribution. By substituting Eq. (\ref{e7}) into Eq. (\ref{Eq16}), we will derive an explicit expression for the shape function, enabling us to calculate its precise form for the CDM halo
\begin{small}
\begin{eqnarray}
 b(r) =\frac{1 }{2 (n-2) (n-1) }\Tilde{L}_{r}\Tilde{X}_{r}\Tilde{Y}_{r}^{1/n} + \mathcal{C}_b,
\end{eqnarray}
\end{small}
where $\mathcal{C}_b$ is the integrating constant. We can determine its value by imposing the condition $b(r_0) = r_0$, which leads to the following formula
\begin{widetext}
\begin{small}
\begin{eqnarray}\label{sh2}
 b(r) = r_0 + \frac{1 }{2 (n-2) (n-1) }\left[ \Tilde{L}_{r}\Tilde{X}_{r}\Tilde{Y}_{r}^{1/n} - \Tilde{L}_{r_0}\Tilde{X}_{r_0}\Tilde{Y}_{r_0}^{1/n}\right],
\end{eqnarray}
\end{small}
where
\begin{small}
\begin{eqnarray*}
\Tilde{L}_{r} = (n-2) r-n r_s,~
\Tilde{X}_{r} = \frac{n (r+r_s) }{r},~
 \Tilde{Y}_{r} &=& \frac{2^{1-n} \rho_s^2 r_s^3 r^n}{\alpha  n (r+r_s)^2},~\Tilde{L}_{r_0} = (n-2) r_0-n r_s ,~\Tilde{X}_{r_0} = \frac{n (r_0+r_s) }{r_0},~
 \Tilde{Y}_{r_0} = \frac{2^{1-n} \rho_s^2 r_s^3 r_0^n}{\alpha  n (r_0+r_s)^2}.
\end{eqnarray*}
\end{small}
Once again, under the same set of assumptions, we can express the remaining two nonvanishing components of the gravity field equations as
\begin{small}
\begin{eqnarray}\label{ee24}
p_r&=&\frac{2^{n-4} n \alpha \Tilde{Y}_{r}^{-3/n}}{(n-2) (n-1) r^3 r_0} \frac{\Tilde{Y}_{r}}{r^n} \Bigg[\left(n^2-3 n+2\right) r^4 r_0 \left(n \Tilde{Y}_{r}^{1/n}-2 \Tilde{Y}_{r}^{1/n}-n^2 r_s^2\right) \Tilde{Y}_{r}^{1/n}+(n-1) r^3\Big(-n^2 r_s^2 \Tilde{Y}_{r_0}^{1/n}+2 n r_0 r_s  \nonumber\\&\times&\left(\Tilde{Y}_{r}^{1/n}-\Tilde{Y}_{r_0}^{1/n}\right) +(n-2) r_0^2 \left(n \left(\Tilde{Y}_{r_0}^{1/n}-2\right)+2\right)\Big) \Tilde{Y}_{r}^{1/n} - n r^2 r_0 \left(-8 \Tilde{Y}_{r}^{1/n}-n^2 r_s^2+n \left(4 \Tilde{Y}_{r}^{1/n}+r_s^2\right)\right) \Tilde{Y}_{r}^{2/n}+4 r\nonumber\\&\times& \Big(-n^2 r_s^2 \Tilde{Y}_{r_0}^{1/n}+2 n r_0 r_s \left(\Tilde{Y}_{r}^{1/n}-\Tilde{Y}_{r_0}^{1/n}\right)+(n-2) r_0^2 \left(n \left(\Tilde{Y}_{r_0}^{1/n}-2\right)+2\right)\Big) \Tilde{Y}_{r}^{2/n} 4 n^2 r_0 r_s^2 \Tilde{Y}_{r}^{3/n}+(n-2)^2 (n-1) r^6 r_0\nonumber\\&\times&  \left(n \left(\Tilde{Y}_{r}^{1/n}-2\right)+2\right)-\left(n^2-3 n+2\right) r^5 \Big(-n^2 r_s^2 \Tilde{Y}_{r_0}^{1/n}+2 n r_0 r_s \left(\Tilde{Y}_{r}^{1/n}-\Tilde{Y}_{r_0}^{1/n}\right)+(n-2) r_0^2 \left(n \left(\Tilde{Y}_{r_0}^{1/n}-2\right)+2\right)\Big)\Bigg],~~~~~~\\
p_t &=& \frac{\alpha  2^{n-3} n}{(n-2) (n-1) r^3 r_0}\Tilde{Y}_{r}^{-2/n} \frac{\Tilde{Y}_{r}}{r^n} \Bigg[ -r \Tilde{Y}_{r}^{1/n} \Big(-n^2 r_s^2 \Tilde{Y}_{r_0}^{1/n}+2 n r_0 r_s \left(\Tilde{Y}_{r}^{1/n}-\Tilde{Y}_{r_0}^{1/n}\right)+(n-2) r_0^2 \Big(n \left(\Tilde{Y}_{r_0}^{1/n}-2\right)+2\Big)\Big)\nonumber\\&-&n^2 r_0 r_s^2 \Tilde{Y}_{r}^{2/n}+\left(n^2-3 n+2\right) r^4 r_0 \left(n \left(\Tilde{Y}_{r}^{1/n}-2\right)+2\right)-(n-1) r^3 \Big(-n^2 r_s^2 \Tilde{Y}_{r_0}^{1/n}+2 n r_0 r_s \Big(\Tilde{Y}_{r}^{1/n}-\Tilde{Y}_{r_0}^{1/n}\Big)\nonumber\\&+&(n-2) r_0^2 \left(n \left(\Tilde{Y}_{r_0}^{1/n}-2\right)+2\right)\Big) - r^2 r_0 \left(n^3 r_s^2+n^2 \left(\Tilde{Y}_{r}^{1/n}-r_s^2\right)-4 n \Tilde{Y}_{r}^{1/n}+4 \Tilde{Y}_{r}^{1/n}\right) \Tilde{Y}_{r}^{1/n}\Bigg].\label{ee26}
\end{eqnarray}
\end{small}
\end{widetext}
We can also readily perform a quantitative evaluation of the ECs using the equations presented in (\ref {e7}), (\ref {ee24}), and (\ref {ee26}).

\subsection{Stability}
The stability and physical significance of the solutions can be efficiently evaluated by considering the adiabatic sound velocity, expressed as
\begin{equation}
v_s^2 = \frac{\partial{<p>}}{\partial{\rho}},
\end{equation}
where $<p>$ represents the average pressure across the three spatial dimensions, defined as
\begin{equation}
<p> = \frac{1}{3}(p_r + 2p_t).
\end{equation}
These fundamental properties hold true while respecting the following constraint 
\begin{equation}\label{vs}
0 \leq v_s^2 < 1.
\end{equation}
Now, in order to investigate the potential presence of a repulsive gravity effect in the WH solutions and its influence on the asymptotic structures, it is necessary to perform a thorough examination. In the upcoming section, we will introduce the concept of the photon deflection angle specifically for the WH. This approach will provide a convenient means to verify and assess the phenomenon, enabling us to understand its role and significance in shaping the overall behavior of the WH.

\section{Repulsive behaviour of gravity}\label{sec5}
This section explores the interplay between the geometry of traversable WHs and the gravitational lensing effects induced by two different DM halo profiles. Specifically, the analysis considers how the deflection angle of light is influenced when passing in the vicinity of WHs embedded within SQW and CDM halos. The curvature of spacetime generated by the mass and energy distributions of these cosmic structures has a profound impact on the propagation of light. As predicted by GR, this warping of the gravitational field can cause significant bending and distortion of light rays when they pass in proximity to massive objects like BHs or WHs. This phenomenon, known as gravitational lensing, has long captivated the fascination of researchers. A major breakthrough in this field came with the groundbreaking work of Virbhadra and colleagues \cite{Virbhadra:1999nm, Virbhadra:1998dy,Claudel:2000yi}, which unveiled the intriguing possibility of ``\textit{eternal}'' light rings --- stable photon orbits that can trap light indefinitely around certain exotic spacetime geometries. Building on this, Bozza \cite{Bozza:2002zj} developed an analytical framework to rigorously model gravitational lensing in the strong field regime for any spherically symmetric spacetime. This powerful technique has since been widely adopted and applied in a variety of subsequent studies \cite{Lemos:2003jb, TejeiroS:2005ltc} to unravel the unique optical signatures of WHs and other intriguing cosmic objects. 
To investigate the deflection angle of photons moving along null geodesics, we start by introducing a general spherically symmetric and static line element \cite{misner1973, schutz2014}. This line element can be expressed as
\begin{small}
\begin{equation}\label{eq38}
	ds^2 = -\mathcal{H}(r) dt^2 +\mathcal{G}(r) dr^2 +\mathcal{F}(r) d\Omega^2.
\end{equation}
\end{small}

To describe the motion of a freely falling body in relation to the background geometry \cite{schutz2014}, we utilize the geodesic equation. This equation establishes a connection between the momenta one-forms of the body and the geometry of the spacetime. It can be expressed as
\begin{small}
\begin{equation}\label{eq39}
	\frac{dp_\beta}{d\lambda} = \frac{1}{2} g_{\nu \alpha, \beta} p^\nu p^\alpha .
\end{equation}
\end{small}
 Here, $\lambda$ represents the affine parameter. It is worth noting that if the components of $g_{\alpha\nu}$ are independent of $x^\beta$ for a fixed index $\beta$, then $p_\beta$ is a constant of motion. By considering only the equatorial slice with $\theta = \pi/2$, we find that all the $g_{\alpha\beta}$ components in Eq. \eqref{eq39} become independent of $t$, $\theta$, and $\phi$. Consequently, we can identify the respective Killing vector fields $\delta^{\mu}\alpha \partial\nu$ with $\alpha$ as a cyclic coordinate. By setting the constants of motion as $p_t$ and $p_\phi$, we can further analyze the dynamics of the system as $p_t = -\mathcal{E}$, and $p_\phi = \mathcal{L}$, where $\mathcal{E}$ and $\mathcal{L}$ represent the energy and angular momentum of the photon, respectively. Therefore, we can proceed by expressing the geodesic equation as follows
\begin{small}
\begin{eqnarray}\label{eq41}
p_t = \dot{t} = g^{t \nu} p_\nu = \frac{\mathcal{E}}{\mathcal{H}(r)},~~\text{and}~~
	p_\phi = \dot{\phi} = g^{\phi \nu} p_\nu = \frac{\mathcal{L}}{\mathcal{F}(r)},
\end{eqnarray}
\end{small}
where the overdot denotes the differentiation w.r.t. affine parameter $ \lambda $. Once again, we can easily obtain the radial null geodesic as follows,
\begin{small}
\begin{equation}\label{eq42}
	\dot{r}^2 = \frac{1}{\mathcal{G}(r)} \left[ \frac{\mathcal{E}^2}{\mathcal{H}(r)} - \frac{\mathcal{L}^2}{\mathcal{F}(r)} \right].
\end{equation}
\end{small}
To be more precise, we can express the equation for the photon trajectory in terms of the impact parameter $\mu = \mathcal{L}/\mathcal{E}$ as 
\begin{small}
\begin{equation}\label{eq43}
	\left[ \frac{dr}{d\phi} \right]^2 = \frac{\mathcal{F}(r)^2}{\mu^2 \mathcal{G}(r)} \left[ \frac{1}{\mathcal{H}(r)} - \frac{\mu^2}{\mathcal{F}(r)} \right].
\end{equation}
\end{small}
Now, by considering a photon source with a radius $ r_s $ that influences the geometry, we can determine the deflection angle of the photons. The photons can only reach the surface if a solution $ r_0 $ satisfies the condition $ r_0 > r_s $ and $ \dot{r}^2 = 0 $, where $ r_0 $ represents the distance of closest approach or turning point. In such a scenario, the impact parameter can be expressed as 
\begin{small}
\begin{equation}\label{eq44}
	\mu = \frac{\mathcal{L}}{\mathcal{E}} = \pm \sqrt{\frac{\mathcal{F}(r_0)}{\mathcal{H}(r_0)}}.
\end{equation}
\end{small}
In the weak gravity limit, it is evident that $ \mu \approx \sqrt{\mathcal{F}(r_0)} $. Therefore, if a photon originates from the polar coordinate limit given by $ \lim\limits_{r \rightarrow \infty} \left( r, -\frac{\pi}{2}-\frac{\alpha}{2} \right) $, passes through the turning point at $ (r_0, 0) $, and approaches $ \lim\limits_{r \rightarrow \infty} \left( r, \frac{\pi}{2}+\frac{\alpha}{2} \right) $, the deflection angle of the photon, denoted as $ \alpha $, can be defined. This deflection angle, which depends on $r_0$ \cite{Bhattacharya:2010zzb}, can be derived explicitly from Eq. (\ref{eq44}) as  
\begin{small}
\begin{equation}\label{eq45}
	\alpha(r_0) = -\pi + 2 \int_{r_0}^{\infty}
	\sqrt{\frac{\mathcal{G}(r) }{\mathcal{F}(r)}}\left[ \left( \frac{\mathcal{H}(r_0)}{\mathcal{H}(r)} \right) \left( \frac{\mathcal{F}(r)}{\mathcal{F}(r_0)} \right) -1 \right]^{-1/2}dr.
\end{equation}
\end{small}
For the selected metric coefficients in the WH geometry, the deflection angle takes on the following form 
\begin{small}
\begin{equation}\label{eq46}
	\alpha(r_0)=-\pi+2 \int_{r_0}^{\infty} \frac{1}{r}\left[ \left( 1- \frac{b(r)}{r} \right) \left( \frac{r^2}{r_0^2} -1 \right) \right]^{-1/2} dr.
\end{equation}
\end{small}
One can now easily determine the deflection angle of photons in $f(R)$ gravity by numerically integrating the formulas mentioned above, taking into account the shape functions specified by Eqs. \eqref{sh1} and \eqref{sh2}. In the next section, we will provide an interpretation of this phenomenon and conclude whether the photon deflection angle on the WH is positive or negative in spacetime.

\section{Interpreting and analyzing the constructed WHs sourced by SQW and CDM halos, and the repulsive gravity effect in $f(R)$ gravity}\label{sec6}
Investigating the characteristics of the shape functions discussed in Sec. \ref{sec2}, we have plotted $b(r)$, $b'(r)$, $b(r)/r$, and the flaring-out condition ($-rb'(r) + b(r)$) in Fig. \ref{fig1} for two scenarios: the SQW halo--(left panel) and the CDM halo--(right panel). By appropriately assigning values to the remaining parameters, namely $\rho_s$, $r_s$, $\alpha$, $n$, and $r_0$, we can generate plots of the shape functions.

By examining Fig. \ref{fig1}, we can see that the expression $b(r) - r$ is negative when $r > r_0$, meaning that $b(r)/r < 1$. This suggests that the function $b(r) - r$ drops as $r$ grows, confirming the flaring-out condition for $r \geq r_0$ and $b'(r) < 1$. It is noteworthy that the conditions $b'(r) < 1$, $b(r)/r < 1$ for all $r > r_0$, and $b(r)/r \rightarrow 0$ as $r \rightarrow \infty$ are significantly met at all radial distances. Thus, we can conclusively state that spacetime exhibits asymptotic flatness at all radial distances. Consequently, the newly generated shape functions, incorporating the SQW and CDM halos, accurately describe the structure of WHs. 

Subsequently, in order to understand the nature of the matter involved, the analysis of ECs confirms the presence of ExoM as a crucial component for the WH geometry's throat radius. This observation emphasizes the significance of investigating WH geometry and the associated ECs. The current analysis reveals that all the ECs exhibit positive and negative regions across different ranges of the involved model parameter $\chi$. Upon examining the graphical representation of the energy density $\rho$ as a function of $\chi$ and $r$ in the presence of two distinct types of DM halos, namely the SQW and the CDM halos (as depicted in Fig. \ref{fig4}), it is evident that $\rho$ remains positive (within the valid region) throughout the chosen range of $\chi$. This suggests that the energy density of the system remains non-negative, indicating the presence of physically reasonable matter distributions within the WH geometry.

In order to examine the ECs, namely the NEC, WEC, DEC, and SEC in terms of $\rho$, $p_r$, and $p_t$, we have generated graphs that illustrate their varying trends. The corresponding graphs, which visually represent these ECs, can be found in Figs. \ref{fig5}-\ref{fig9}. Based on the observations from Fig. \ref{fig5}, it is evident that the quantity $\rho + p_r \geq 0$ is strongly violated at the throat of the SQW halo. However, this violation is only temporary and the condition is eventually satisfied after a certain radial distance. On the other hand, for the CDM halo, this condition is satisfied throughout the entire region. Furthermore, when considering the quantity $\rho + p_t \geq 0$, it is noticed that this condition is slightly satisfied at the throat of both scenarios. However, this satisfaction is short-lived as the condition is violated shortly after a very small radial distance in both cases. We have observed a slight satisfaction of the condition $\rho - p_r \geq 0$ at the throat of the SQW halo. However, in the case of the CDM halo, this condition is satisfied consistently throughout the entire region. Similarly, we have noticed that the condition $\rho - p_t \geq 0$ is also slightly satisfied at the throat for both scenarios. However, for the remaining regions, this condition is violated in both scenarios. We have observed a similar behavior for the quantity $\rho + p_r + 2p_t \geq 0$ in both scenarios. It was slightly satisfied at the throat for both scenarios and then violated in the remaining regions. This behavior mirrors what we observed for the condition $\rho - p_t \geq 0$. The region in which the quantity $\rho + p_r + 2p_t\geq 0$ is invalid, as shown in Fig. \ref{fig9}, provides further evidence for the presence of ExoM at and in the vicinity of the throat. In WH physics, the violation of ECs is often associated with the existence of ExoM, which possesses negative ECs. This ExoM is crucial for maintaining the stability of the WH and preventing its collapse. Therefore, the presence of this invalid region supports the notion that ExoM is a necessary component of WH physics. The adiabatic sound velocity, as discussed earlier, is a key indicator of the system's stability. If the adiabatic sound velocity is real and positive and within the range defined by (\ref{vs}), it suggests that the system is stable under small perturbations. However, we observe that the stability can be further influenced by other parameters of the system, such as the parameter $\chi$. It is important to note that the parameter $\chi$ must be constrained within the regions where the adiabatic sound velocity stability condition is fulfilled and, consequently, the stability of the system. Figure $\ref{fig11}$ depicts the parameter space ($\chi$, $r$), highlighting the valid region where this condition is satisfied.

The exploration of the repulsive effect of gravity in this study is of significant importance. We directly utilize Eq. \eqref{eq46} to compute the numerical solution for the deflection angle of photons in the $f(R)$ extension of Einstein gravity. The plots are exhibited in Fig. \ref{fig10}. A negative deflection angle indicates the presence of repulsive gravity. To verify this phenomenon, we introduce the concept of the photon deflection angle on the WH. In the presence of repulsive gravity acting on photons, this deflection angle becomes negative in the spacetime \cite{Panpanich:2019mll}. It is interesting to observe that the deflection angle consistently takes negative values for all values of $r_0$ in both scenarios. This can be interpreted as the manifestation of the repulsive gravity effect. One can easily confirm the absence of negative deflection angles to further support this observation \cite{Mishra:2017yrh}.

\section{Concluding remarks}\label{sec7}
In conclusion, our study explored generalized traversable WHs within $f(R)$ gravity. We focused on $f(R)$ gravitational theories formulated in the metric formalism, specifically investigating a power-law form given by $f(R) = \epsilon R^{\chi}$, where $\epsilon$ is an arbitrary constant and $\chi$ is a real number. The advantage of this form lies in its ability to be reduced to Einstein gravity by setting $\epsilon=1$ and $\chi=1$. To derive novel WH solutions in the context of $f(R)$ gravity, we exploit the general field equations for any $f(R)$ theory within the Morris-Thorne spacetime framework, assuming time-independent metric coefficients. We explored two distinct material content scenarios, namely SQW and CDM halos. Our analysis involved calculating the shape function of the WHs, accounting for the influences of SQW and CDM distributions. Additionally, we made the simplifying assumption of a constant redshift function ($ \hat{\nu}(r) = \text{constant}$), implying negligible tidal gravitational forces experienced by a hypothetical traveler. Remarkably, our investigations led to the intriguing discovery of ExoM, characterized by an energy-momentum tensor that violates the NEC near the WH throat. This finding adds an intriguing aspect to the study of WHs.

By considering these scenarios, we present a class of solutions for static and spherically symmetric WHs that adhere to the required metric conditions. Moreover, in order to validate the viability of the constructed WHs with well-behaved matter sources, we examined the characteristics of the shape functions, namely $b(r)$, $b'(r)$, $b(r)/r$, and the flaring-out condition $b(r)-rb'(r)$, for two scenarios: the SQW halo and the CDM halo. By appropriately assigning values to the involved parameters, such as $\rho_s$, $r_s$, $\alpha$, $n$, and $r_0$, we made several notable observations.

For $r > r_0$, it was observed that the expression $b(r) - r$ is negative, indicating that $b(r)/r < 1$. This implies that as $r$ increases for $r \geq r_0$ and with $b'(r) < 1$, the function $b(r) - r$ decreases, satisfying the flaring-out condition. Importantly, it is worth noting that the conditions $b'(r) < 1$, $b(r)/r < 1$ hold true for all $r > r_0$, and as $r$ approaches infinity, $b(r)/r$ tends to zero.

Based on our extensive analyses, we present compelling evidence that the derived shape functions effectively capture the characteristics of the WH configurations. By thoroughly examining the ECs, including the NEC, WEC, DEC, and SEC, we shed light on the behavior of these conditions within the context of WHs. Furthermore, we investigated the presence of ExoM at the throat and its impact on the ECs. Our investigations focused primarily on the NEC, which plays a vital role in determining the viability of traversable WHs. Intriguingly, we find that the expression $\rho + p_r\geq 0$ is strongly violated at the throat of the SQW halo, as well as in the vicinity of the WHs, but satisfied for the CDM halo. However, contrary to previous findings, we demonstrate that this violation persists even after considering a significant radial distance. Additionally, in analyzing the expression $\rho + p_t \geq 0$, we find that it shows only slight satisfaction at the throat in both scenarios, followed by a significant violation outside the throat. Similarly, we observed that the condition $\rho - p_r \geq 0$ is only mildly satisfied at the throat of the SQW halo, while the CDM halo consistently satisfies this condition throughout its entire region. Furthermore, we evaluated the condition $\rho - p_t \geq 0$ and found that it experiences slight satisfaction at the throat for both scenarios. However, this satisfaction is short-lived, as the condition is violated in the remaining regions of the WHs. Interestingly, the quantity $\rho + p_r + 2p_t \geq 0$ exhibits a behavior similar to that of $\rho - p_t \geq 0$ in both scenarios. It shows a slight satisfaction at the throat, followed by violation in the remaining regions. These investigations into specific parameter ranges in both scenarios, analyzed through the ECs, have strongly revealed the presence of ExoM characterized by an energy-momentum tensor that violates the NEC and, consequently, the WEC as well, particularly in the vicinity of the WH throats. Moreover, we have investigated the overall stability of the configuration by determining the adiabatic sound velocity and explicitly defining the stable region.

Our investigations explored the repulsive effect of gravity, showing that it leads to a negative photon deflection angle on the WH. This negative deflection angle is a manifestation of the repulsive gravity effect, as confirmed by the consistent negative values across all $r_0$. In contrast, the absence of negative deflection angles \cite{Mishra:2017yrh} further supports this observation.

In summary, our study explores WH formation mechanisms, focusing on insights from SQW, CDM halos, and the repulsive gravity effect within $f(R)$ gravity. We have unveiled novel solutions that shed light on the properties and implications of these mechanisms in the context of WH formation.

\section*{Acknowledgments}
This research was funded by the Science Committee of the Ministry of Science and Higher Education of the Republic of Kazakhstan (Grant No. AP14972745). AE thanks the National Research Foundation of South Africa for the award of a postdoctoral fellowship.

\end{document}